\documentclass[aps,prb,twocolumn,floatfix,amsmath,amssymb]{revtex4}
\usepackage[final]{graphicx}
\usepackage{bm}
\usepackage{float}
\usepackage{afterpage}

\begin{document}
\title{Flat bands with non-trivial topology in three dimensions}
\author{C. Weeks}
\affiliation{Department of Physics and Astronomy, University of British Columbia, Vancouver, BC, Canada V6T 1Z1}
\author{M. Franz}
\affiliation{Department of Physics and Astronomy, University of British 
Columbia, Vancouver, BC, Canada V6T 1Z1}
\begin{abstract}
We construct a simple model for electrons in a three-dimensional crystal where a combination of short-range hopping and spin-orbit coupling results in nearly flat bands characterized by a non-trivial $\mathbb{Z}_2$ topological index. The flat band is separated from other bands by a bandgap $\Delta$ much larger than the bandwidth $W$. When the flat band is partially filled we show that the system remains non-magnetic for a significant range of repulsive interactions. In this regime we  conjecture that the true many-body ground state may become a three-dimensional fractional topological insulator.
 \end{abstract}

\maketitle

{\em Introduction. --}
When single-particle states of bosons or fermions comprise a flat band, the effect of interactions is generically non-perturbative and can lead to new states of quantum  matter. For 2D electrons in a strong perpendicular magnetic field the flatness of Landau levels combined with non-zero Chern numbers leads to the spectacular fractional quantum Hall (FQH)  effects\cite{tsui1,laughlin1} and other exotic states such as the Wigner crystal.\cite{lam1} Recently, models for spinless fermions moving in two-dimensional lattices have been constructed that exhibit nearly-flat bands with non-zero Chern numbers, but remarkably in {\em zero} external magnetic field.\cite{tang,sun,neupert,hu} It has been proposed, on the basis of heuristic arguments\cite{tang, sun} and numerical simulations,\cite{neupert,sheng1,regnault1} that at partial filling repulsive interactions may drive these systems into the zero-field FQH states.
Although it is not clear apriori how to experimentally realize the above situation in a physical system, the prospect of engineering systems that can support fractional excitations without involving a magnetic field has led
to significant interest recently in flat or nearly flat bands in various lattice models. 

In this Communication we ask whether it is possible to achieve a flat band with non-trivial topology in a system of electrons in three spatial dimensions. It is well known that the concept of the first Chern number (and the FQH physics) does not generalize to three dimensions. However, recent advances in the theory of time-reversal $({\cal T})$ invariant band insulators have established a uniquely three-dimensional $\mathbb{Z}_2$-valued topological invariant\cite{moore1,hasan1,qi1} called $\nu_0$  which, as we argue below, can play a similar role. Taking a specific example of electrons moving in the 3D edge-centered cubic (perovskite) lattice  we show that a suitable combination of short-range hoppings and spin-orbit coupling (SOC)  terms can give rise to a nearly flat band with a non-trivial $\mathbb{Z}_2$ index $\nu_0=1$. With the chemical potential inside the bandgap such crystal would be a strong topological insulator with an odd number of topologically protected gapless states associated with all of  its surfaces. At partial filling and in the presence of repulsive interactions we argue that the system may become a three-dimensional `fractional' topological insulator.\cite{pesin1,maciejko1,swingle1,qi2} 

Many lattice models are known to support completely flat bands. In 2D these include $p$-orbitals in the honeycomb lattice,\cite{wu1} as well as $s$-orbitals in the kagome, dice and Lieb lattices.\cite{bergman1,guo1,weeks1} 3D examples include the pyrochlore and the perovskite lattice.\cite{bergman1,guo2,weeks1}
In their simplest form however the above tight-binding models do not constitute a platform suitable for the study of exotic correlated phases for at least two reasons. First, in all cases the flat bands touch other dispersing bands, usually in a quadratic fashion. The interactions are thus bound to mix states from adjacent bands and this is clearly detrimental to the emergence of correlated phases. Second, even when the Hamiltonian is modified  to open a gap, e.g.\ by adding SOC as in Ref.\ \onlinecite{weeks1}, the resulting flat (or nearly-flat) bands are in all known cases topologically trivial. In such topologically trivial flat bands the eigenstates can be chosen as exponentially localized around lattice sites and, at partial filling, interactions will typically select a crystalline ground state with broken translational symmetry and not an exotic featureless liquid that underlies the FQH effect. As already argued in Refs.\ \onlinecite{neupert, sun, tang, weeks1} a natural place to seek such exotic phases is in a flat band that is {\em topologically non-trivial}. 

It is well known that in 2D a non-zero Chern number represents a topological obstruction to the formation of exponentially localized Wannier states.\cite{thouless1} It is precisely this obstruction that tilts the balance in favor of FQH states in partially filled Landau levels (although Wigner crystal phases are still known to arise at very low filling fractions).
For $\mathbb{Z}_2$-odd phases of ${\cal T}$-invariant band insulators in 2D and 3D a similar topological obstruction exists\cite{vanderbilt1} if one insists on Wannier states that respect ${\cal T}$. This consideration suggests that interacting electrons partially filling a flat band with a non-trivial $\mathbb{Z}_2$ index in a 3D crystal could either (i) spontaneously break ${\cal T}$ and  form a Wigner crystal, or (ii) remain ${\cal T}$-invariant and form one of the proposed  3D fractional topological insulators characterized either by spin-charge separation\cite{pesin1} or  by fractional magneto-electric effect  controlled by fractional values of the axion parameter $\theta$ accompanied by ground state degeneracy. \cite{maciejko1,swingle1,qi2}

In the following we lay groundwork for future investigations of these 3D exotic phases by constructing a simple tight-binding model whose spectrum has a flat band characterized by $\nu_0=1$ and is separated from other bands by a large gap. We remark that constructing a flat band with non-trivial topology in a 3D crystal is mathematically more constrained problem than in 2D and it is by no means obvious that this can be achieved with short-range hoppings only.

\begin{figure}[t]
\begin{center}
\includegraphics[scale=0.27]{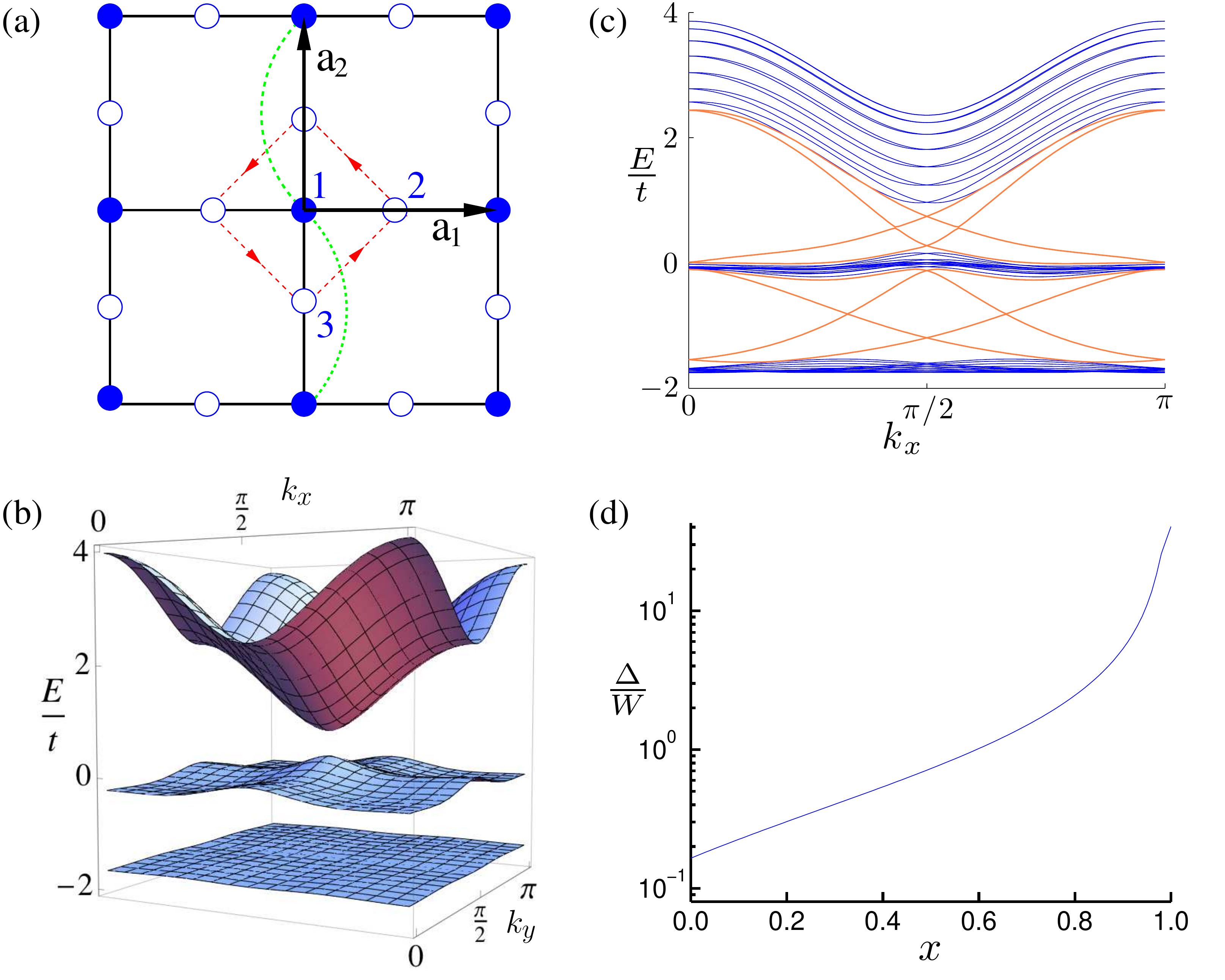}
\caption{ (Color online)(a) Lieb lattice showing 3-site basis in unit cell and  basis vectors $\mathbf{a}_1$ and $\mathbf{a}_2$.  Second and third neighbor hoppings are indicated by red and green dotted lines, respectively. (b) Tight-binding dispersion for ${\cal H}_{\bf k}$ after tuning parameters to achieve maximal figure of merit $\Delta/W$. For clarity the bulk energy bands are plotted over the shifted  Brillouin zone $0\leq k_{x}\leq\pi/a$, $0\leq k_{y}\leq\pi/a$ where $a$ represents the length of the nearest neighbor bond. (c) Bandstructure for strip of $N_{y}=10$ unit cells with open boundary conditions along the $y$ direction and infinite along $x$ with the same parameters as in (b). (d) Semi-Log plot showing the evolution of the gap as one moves between a known topological insulating phase and the final set of finely tuned parameters.} \label{unit_cell}
\end{center}
\end{figure}

\begin{figure}[t]
\begin{center}
\includegraphics[scale=0.31]{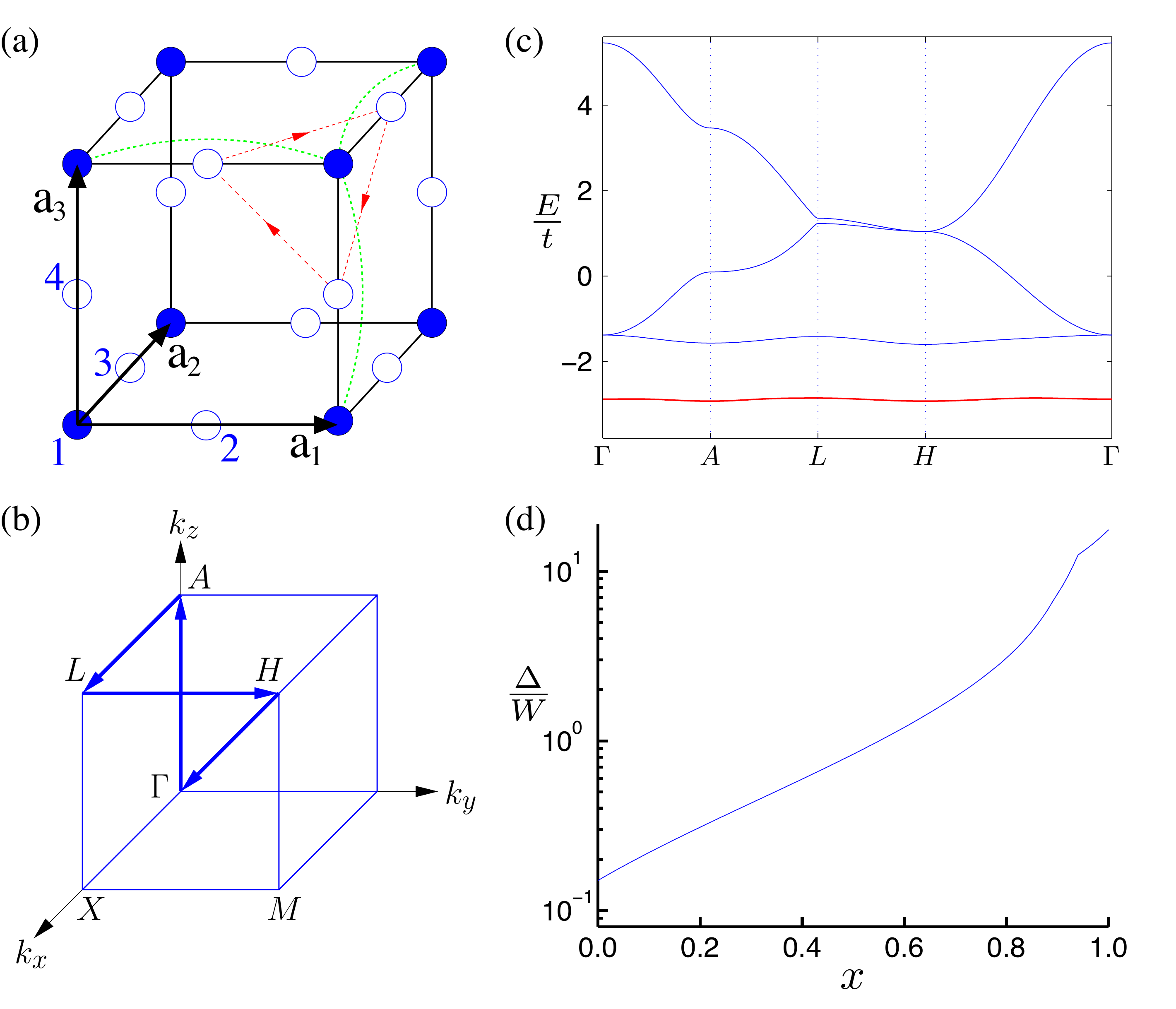}
\caption{(Color online) (a) Perovskite lattice showing 4 sites in unit cell along with basis vectors. Second and third neighbor hoppings are indicated by red and green dotted lines, respectively. (b) High symmetry points in the Brillouin zone. (c) Bandstructure inside bulk along path of high symmetry for ${\cal H}_{\bf k}$ with finely tuned parameter set. (d) Semi-Log plot showing the evolution of the gap as one moves between a known topological insulating phase and the final set of finely tuned parameters. } \label{perovskite}
\end{center}
\end{figure}

{\em The model. --} As a first step in the program outlined above we study a simple tight-binding model with intrinsic SOC terms on the 2D Lieb lattice, seen in Fig.~\ref{unit_cell}a, and the 3D peroviskite lattice seen in Fig.~\ref{perovskite}a. The relevant Hamiltonian reads
\begin{equation}
\label{h0}
\begin{split}
H&=-\sum_{\langle ij\rangle\alpha}t_{ij}c_{i\alpha}^{\dag}c_{j\alpha} -\delta\sum_{i\in {\rm corner}}c_{i\alpha}^{\dag}c_{i\alpha} 
\\&+i{\lambda}\sum_{\langle\langle ij\rangle\rangle\alpha\beta}
 ({\bf d}_{ij}^1\times {\bf d}_{ij}^2)\cdot{\bm \sigma}_{\alpha\beta} c^\dagger_{i\alpha} c_{j\beta},
 \end{split}
\end{equation} 
where $c^{\dagger}_{i\alpha}$ creates an electron of spin $\alpha$ on site $i$ of  the lattice, $t_{ij}$  are the hopping amplitudes which we take equal to $t_{1}$, $t_{2}$,and  $t_{3}$ for the first, second and third nearest neighbor sites respectively, and $\delta$ is an onsite energy for the corner sites (indicated as filled circles in Figs.\ 1a and 2a). Lastly, $\lambda$ represents the amplitude for the next-nearest neighbour SOC
where $\mathbf{{d}}^{1}_{ij}$ and
$\mathbf{{d}}^{2}_{ij}$ are the two unit vectors along the nearest neighbour bonds connecting site $i$
to its next-nearest neighbour $j$ and ${\bm \sigma}$ is the vector of Pauli
spin matrices.

In Ref.~\onlinecite{weeks1}, we showed that for nearest neighbor hopping ($t_{1}$ only) and non-zero $\lambda$ both Lieb and perovskite lattices became topological insulators possessing a nontrivial  $\mathbb{Z}_2$ invariant. In both cases the bands carrying non-trivial invariants were strongly dispersing. From here then, our goal  is to flatten out these bands by tuning the hopping strengths $t_{2}$, $t_{3}$, $\lambda$ and the onsite energy $\delta$ without leaving the topological phase. Also, we want the flat band to be separated from all other bands by a large gap $\Delta$ so that interactions do not mix states from different bands. The relevant figure of merit,\cite{tang,sun,neupert} then, is the ratio of the flat-band width $W$ to the bandgap $\Delta$.

{\em Lieb lattice. --} 
As a warm-up exercise we first consider the valence band in the 2D Lieb lattice. Going over to momentum space and diagonalizing the Hamiltonian ${\cal H}_{\bf k}$ that follows from Eq.\ (\ref{h0}), the result of the tuning procedure can be seen in Fig. \ref{unit_cell}b.

The procedure used for the tuning was the simplest available, namely introducing the parameters one by one and modifying the values by hand until the band was as flat as we could achieve. For the following set of values: $t_{1}=1$, $t_{2}=-0.17$, $t_{3}=-0.114$, $\lambda=0.303$ and $\delta=-0.725$, the ratio of the bandwidth to the bandgap, $\Delta/W\approx40.45$. A systematic numerical minimization technique could probably improve upon our result, but comparing to previous results on the kagome, honeycomb, square, checkerboard and ruby lattices, \cite{neupert, sun, tang, hu} this value is already quite respectable.

To show that the system has not migrated away from the topological phase during the parameter tuning stage, we first solve the system in a strip geometry numerically using the same parameters as above. This verifies that the spin-filtered gapless edge states, seen in Fig. \ref{unit_cell}c, indeed persist. 
We also consider the Hamiltonian
\begin{equation}\label{hh}
 H'\left(x\right)=(1-x)H_0+xH,
\end{equation}
where $H_0$ includes only $t_{1}$ and $\lambda$ and is known to support the topological phase.\cite{weeks1} The result for $\Delta/W$, starting at $\lambda=0.1$ ($t_{1}=1$) then adiabatically tuning $x$ from 0 to 1, for the same final set of parameters in $H$, can be seen in Fig. \ref{unit_cell}(d). 
The ratio increases from its initial value $\Delta/W\approx 0.17$ to its largest value 40.45 at $x=1$, without closing the gap. The system is seen to remain in the topological phase.

What can one say about the fate of the ground state in the presence of repulsive interactions when the flat band is partially filled? Close to half filling, since the spin up and down electrons are decoupled, one expects the Stoner instability towards the ferromagnetic state. If we parametrize the interaction by the usual on-site Hubbard $U$ then a gap $\sim U$ should open 
between the majority and minority bands when $W<U<\Delta$. For large enough $U$ one can focus on the spin-polarized electrons in the majority band. In the presence of sufficiently strong residual interactions (e.g. nearest-neighbor repulsion $V$) and at fractional filling, this problem becomes similar to that considered in Refs.\ [\onlinecite{neupert, sun, tang}] with spinless fermions. Analogous arguments then suggest the emergence of FQH states under favorable conditions. Another possibility is the formation of a ${\cal T}$-invariant fractional topological insulator\cite{levin} which can be pictured as two decoupled FQH states for the two spin species. Such a state might be favored if $U\ll V$.

{\em Perovskite Lattice. --} We now approach the main objective of this work: constructing flat bands with a non-trivial $\mathbb{Z}_2$ index in a 3D lattice. 
Our starting point is again the Hamiltonian given in Eq. (\ref{h0}) but now on the perovskite lattice, shown in Fig. \ref{perovskite}(a). 
In Ref.~\onlinecite{weeks1} we showed that for nearest-neighbor hopping and $\lambda>0$ the system becomes a $(1;111)$ strong topological
insulator when the lowest doubly degenerate band is filled. In an analogous fashion to the two dimensional case then, we aim to flatten out the lowest energy band by means of including additional parameters $t_2$, $t_3$ and $\delta$.  The bandstructure for the set of parameters that maximize the bandgap to bandwidth ratio is shown in Fig. \ref{perovskite}c, over a path of high symmetry in the Brillouin zone illustrated in Fig. \ref{perovskite}b. The following set of values:  $t_{1}=1$, $t_{2}=-0.416$, $t_{3}=-0.047$, $\lambda=-0.331$ and $\delta=1.318$, yielded the ratio $\Delta/W\approx17.56$. We remark that the ratio was calculated over the entire Brillouin zone and in Fig. \ref{perovskite}c we have simply chosen one possible path here to illustrate the result. Including additional parameters in $H$, such as longer range hoppings, could no doubt further improve the above figure of merit but we do not pursue this here. 

To ensure that the system remains in the topological phase, we adiabatically tune the parameter $x$ from 0 to 1 in the Hamiltonian $H'(x)$ defined in Eq.\ (\ref{hh}). The result of this procedure can be seen in Fig. \ref{perovskite}d. The ratio, again, remains finite across the entire range from its initial value $\Delta/W\approx 0.15$ to its largest value 17.56 at $x=1$. We conclude that the flat-band carries $\mathbb{Z}_2$ index (1;111).

Now imagine that the (doubly degenerate) flat band is partially filled and consider the effect of repulsive interactions. Unlike the 2D case discussed above where the residual U(1) spin symmetry is preserved despite the presence of SOC, in the 3D strong topological insulator the SU(2) spin symmetry is completely broken. In this situation it is not clear what the leading instability might be. In a similar flat-band setting describing Ir-based pyrochlore Y$_2$Ir$_2$O$_7$ Pesin and Balents\cite{pesin1} argued for an exotic ${\cal T}$-invariant  spin-charge separated topological Mott insulator while others found more conventional magnetic phases.\cite{savrasov1, kim1} Below, we investigate the magnetic instabilities of our model in the simplest case with on-site repulsion and at exact half filling of the flat bands. We show that the system remains non-magnetic when interaction strength is below the critical value which is large compared to the bandwidth $W$. In this regime, therefore, the true many-body ground state could become a ${\cal T}$-invariant fractional topological insulator.
\begin{figure}[b]
\begin{center}
\includegraphics[scale=0.223]{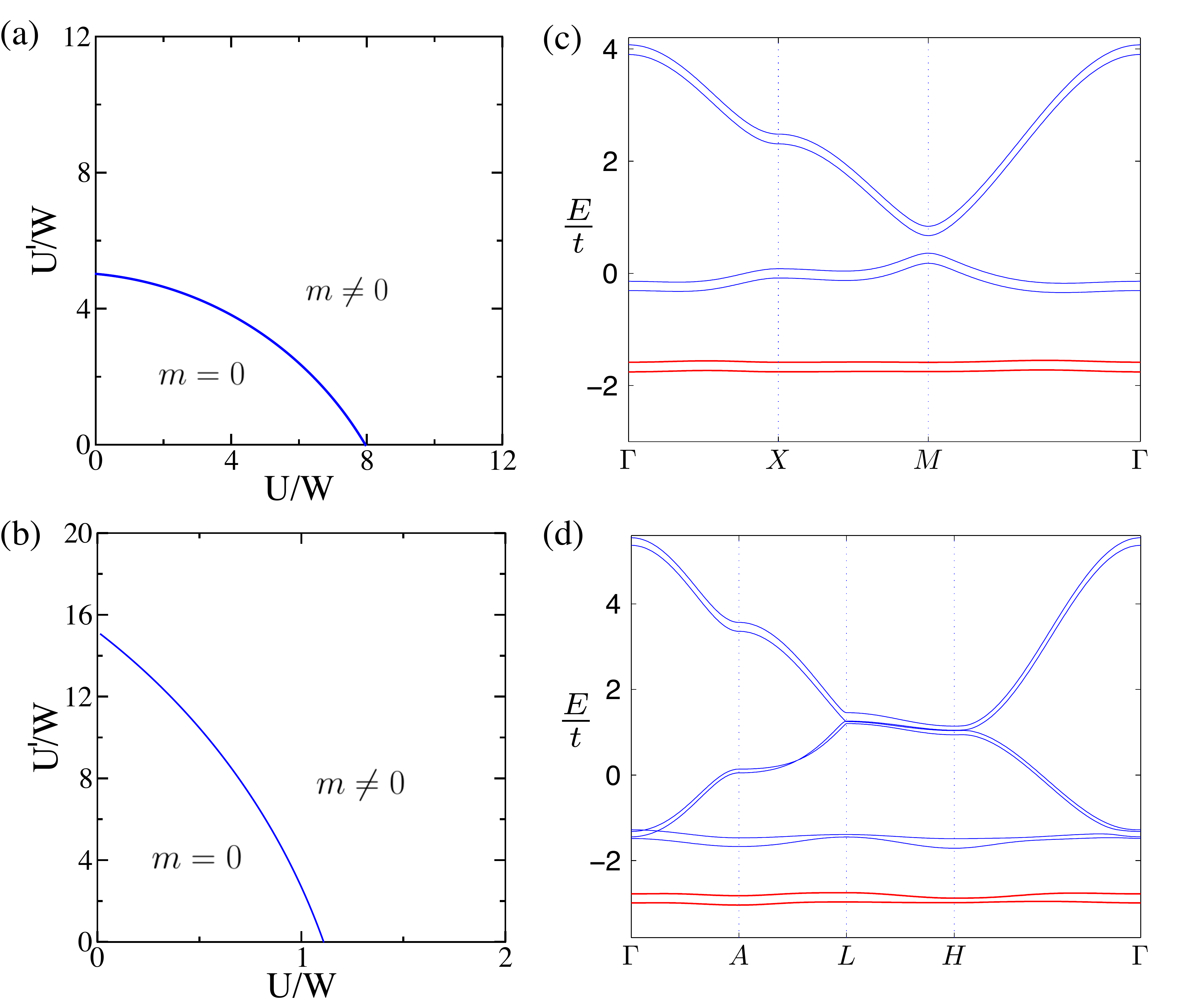}
\caption{ (Color online)(a) Phase diagram for Lieb lattice ($W\approx0.034$), and (b) perovskite lattice ($W\approx0.072$). Critical values $U_c$, $U_c'$ for the onset of magnetic order reflect the charge distribution on the basis sites. The latter is approximately uniform for the Lieb lattice implying  $U_c\approx U_c'$. On perovskite lattice most of the charge density is on site 1 causing $U_c\ll U_c'$.  (c) Band structure for Lieb lattice with $U/W=U'/W=15$, $m^{(1)}\approx0.35$ and $m^{(2)}=m^{(3)}\approx0.32$, and   (d) perovskite lattice with $U/W=5$, $U'/W=25$, $m^{(1)}\approx0.62$, $m^{(2)}=m^{(3)}\approx0.11$ and $m^{(4)}\approx0.05$.} \label{magnetic}
\end{center}
\end{figure} 

{\em Magnetic Instabilities. --}  We extend the Hamiltonian (1) above by including a Hubbard term
\begin{equation}
\label{hubbard}
H_{\rm int}= U\sum_{i}n_{i\uparrow}^{\left(1\right)}n_{i\downarrow}^{\left(1\right)}+U'\sum_{i,l>1}n_{i\uparrow}^{\left(l\right)}n_{i\downarrow}^{\left(l\right)}\end{equation} 
with the goal of mapping out the magnetic phases as a function of $U$ and $U'$ by means of a standard mean-field calculation. The superscript $l$ denotes the basis sites in the lattice, Figs.\ 1a and 2a.  We use the following decoupling
 \begin{equation}
 n_{i\uparrow}n_{i\downarrow} \rightarrow n_{i\uparrow}\langle n_{i\downarrow} \rangle+n_{i\downarrow}\langle n_{i\uparrow} \rangle- \langle n_{i\uparrow} \rangle\langle n_{i\downarrow} \rangle 
\end{equation}
and definine the magnetization $m_{i}^{\left(l\right)}=\langle n_{i\uparrow}^{\left(l\right)} \rangle-\langle n_{i\downarrow}^{\left(l\right)} \rangle$ and occupation number $\bar{n}_{i}^{\left(l\right)}=\langle n_{i\uparrow}^{\left(l\right)} \rangle+\langle n_{i\downarrow}^{\left(l\right)} \rangle$.
 Furthermore, we focus on uniform magnetic phases, such that $m_{i}^{\left(l\right)}=m^{\left(l\right)}$ independent of the site index $i$. 

To map out the phase diagram, we  minimize  the ground state energy with respect to $m^{\left(l\right)}$. The result for half filling of the flat bands can be seen in Figures~\ref{magnetic}a and \ref{magnetic}b, for the Lieb and perovskite lattices respectively. Here $m=\sum_l m^{\left(l\right)}$ is the total magnetization per unit cell. The key observation to be made here is that both systems remain non-magnetic over a significant range of interaction strengths.


{\em Conclusions. --}
We have established that it is possible to engineer nearly flat bands characterized by non-trivial $\mathbb{Z}_2$ topological invariants in both the 2D Lieb lattice and its three dimensional counterpart, the edge centered cubic (perovskite) lattice.  With our 2D example we have simply added to the growing number of lattices capable of producing this behavior, whereas the 3D result is to the best of our knowledge completely novel. What makes these findings potentially broadly relevant is that they provide a concrete framework for  future studies addressing the nature of the many-body ground state that occurs in the presence of repulsive interactions and at partial filling. 

Unlike the 2D case where the Laughlin liquid\cite{laughlin1} furnishes a well established paradigm for the topologically ordered correlated ground state, in 3D such a template is presently lacking. Therefore,
understanding the fate of the ground state of electrons in the topologically non-trivial 3D flat band in the presence of strong interactions beyond the simple mean-field analysis is clearly a problem outside the scope of this study. In this work we made a first step in this direction by demonstrating, within the mean-field approximation, that the ground state remains non-magnetic over a significant portion of the $U$-$U'$ phase diagram (Fig 3b). Since mean-field calculations tend to overestimate the stability of ordered phases the actual non-magnetic region will likely be somewhat larger and can in principle host a ${\cal T}$-invariant fractional topological insulator.
Our result in 3D thus provides a concrete foundation for addressing these interesting questions using the suite of analytical approaches or numerical many-body techniques and we hope that it might inspire further studies.

Finally we remark that there exist many perovskites in nature.  Our theoretical results demonstrate that a near-ideal situation (in terms of strong interaction physics) can arise in a very simple model in this family of lattices. Also alluded to previously\cite{weeks1}  was the possibility of engineering a relevant 2D system by modulating a two-dimensional electron gas with a periodic potential having Lieb lattice symmetry, in an analogous fashion to the `artificial graphene' created recently.\cite{west1} Another possibility lies with cold fermionic or bosonic atoms in optical lattices as discussed in Refs.\ [\onlinecite{DasSarma, Zhang3}], with a detailed study on the Lieb lattice provided in Ref.~ [\onlinecite{goldman}].

\emph{Acknowledgment.}---The authors have benefited from discussions with D. Bercioux, C. Chamon and J.E. Moore. This work was supported by NSERC and CIfAR.


\end{document}